\documentclass[a4paper]{cas-dc}
\usepackage{balance}
\usepackage[numbers,sort&compress]{natbib}
\usepackage{amsmath}
\definecolor{customcolor}{RGB}{0, 102, 204} 
\usepackage[numbers]{natbib}
\usepackage{enumitem}
\usepackage{graphicx}
\usepackage{float}
\usepackage{caption}
\usepackage{subcaption}
\usepackage{epstopdf}
\usepackage{threeparttable}
\usepackage{tabularx}
\usepackage{booktabs}
\usepackage{caption}
\usepackage{multicol}
\usepackage{hyperref}
\usepackage{url}
\usepackage{colortbl}
\hypersetup{
    colorlinks=true,
    linkcolor=customcolor, 
    citecolor=customcolor, 
    filecolor=customcolor, 
    urlcolor=black   
}
\usepackage{xcolor,soul}
\usepackage{adjustbox}
\usepackage{algorithmic}
\usepackage{algorithm}
\usepackage[none]{hyphenat}

\definecolor{LightGray}{rgb}{1,1,1}

\def\tsc#1{\csdef{#1}{\textsc{\lowercase{#1}}\xspace}}
\tsc{WGM}
\tsc{QE}

\begin{document}
\let\WriteBookmarks\relax
\def\floatpagepagefraction{1}
\def\textpagefraction{.001}
\let\printorcid\relax 


\shorttitle{}    

\shortauthors{Yun Xiao et al.}

\title[mode = title]{Multi-Source EEG Emotion Recognition via Dynamic Contrastive Domain Adaptation}  

\author[1]{Yun Xiao}
\ead{yxiao@nwu.edu.cn}

\author[1]{Yimeng Zhang}
\ead{zhangyimeng@stumail.nwu.edu.cn}

\author[3]{Xiaopeng Peng}
\ead{xxp4248@rit.edu}

\author[1]{Shuzheng Han}
\ead{hanshuzheng@stumail.nwu.edu.cn}

\author[2]{Xia Zheng}\corref{cor1}
\ead{zhengxia@zju.edu.cn}

\author[1]{Dingyi Fang}
\ead{dyf@nwu.edu.cn}

\author[1]{Xiaojiang Chen}
\ead{xjchen@nwu.edu.cn}

\cortext[cor1]{Corresponding author.}
\address[1]{School of Information Science and Technology, Northwest University, Xi'an Shaanxi 710069, China.}
\address[2]{School of Art and Archaeology, Zhejiang University, Hangzhou Zhejiang 310018, China.}
\address[3]{Rochester Institute of Technology, Rochester NY 14623, USA.}

\begin{abstract}
Electroencephalography (EEG) provides reliable indications of human cognition and mental states. Accurate emotion recognition from EEG remains challenging due to signal variations among individuals and across measurement sessions. We introduce a multi-source dynamic contrastive domain adaptation method (MS-DCDA) based on differential entropy (DE) features, in which coarse-grained inter-domain and fine-grained intra-class adaptations are modeled through a multi-branch contrastive neural network and contrastive sub-domain discrepancy learning. Leveraging domain knowledge from each individual source and a complementary source ensemble, our model uses dynamically weighted learning to achieve an optimal tradeoff between domain transferability and discriminability. The proposed MS-DCDA model was evaluated using the SEED and SEED-IV datasets, achieving respectively the highest mean accuracies of $90.84\%$ and $78.49\%$ in cross-subject experiments as well as $95.82\%$ and $82.25\%$ in cross-session experiments. Our model outperforms several alternative domain adaptation methods in recognition accuracy, inter-class margin, and intra-class compactness. Our study also suggests greater emotional sensitivity in the frontal and parietal brain lobes, providing insights for mental health interventions, personalized medicine, and preventive strategies. 
\end{abstract}

\maketitle 

\section{Introduction}
\label{sec:introduction}
Emotions are fundamental to human experience and play a significant role in well-being, behavior, social interaction, decision-making, and cognitive function \citep{van2010emerging, quaglia2015mindful, park2023emotional, rolls2023emotion}. Recent advances in electrode technology \citep{paranjape2001electroencephalogram, thatcher2010validity, petrossian2023advances} and machine learning have led to improved electroencephalogram (EEG) analysis for emotional recognition \citep{alarcao2017emotions, li2022eeg, jafari2023emotion}. However, the inherent non-stationarity of EEG signals presents challenges in generalizing an emotion recognition method for accurate predictions between individuals or over time \citep{wu2020transfer, wan2021review, weng2024self}. 

\begin{figure}[!t]
\centering
\includegraphics[width=2.95in]{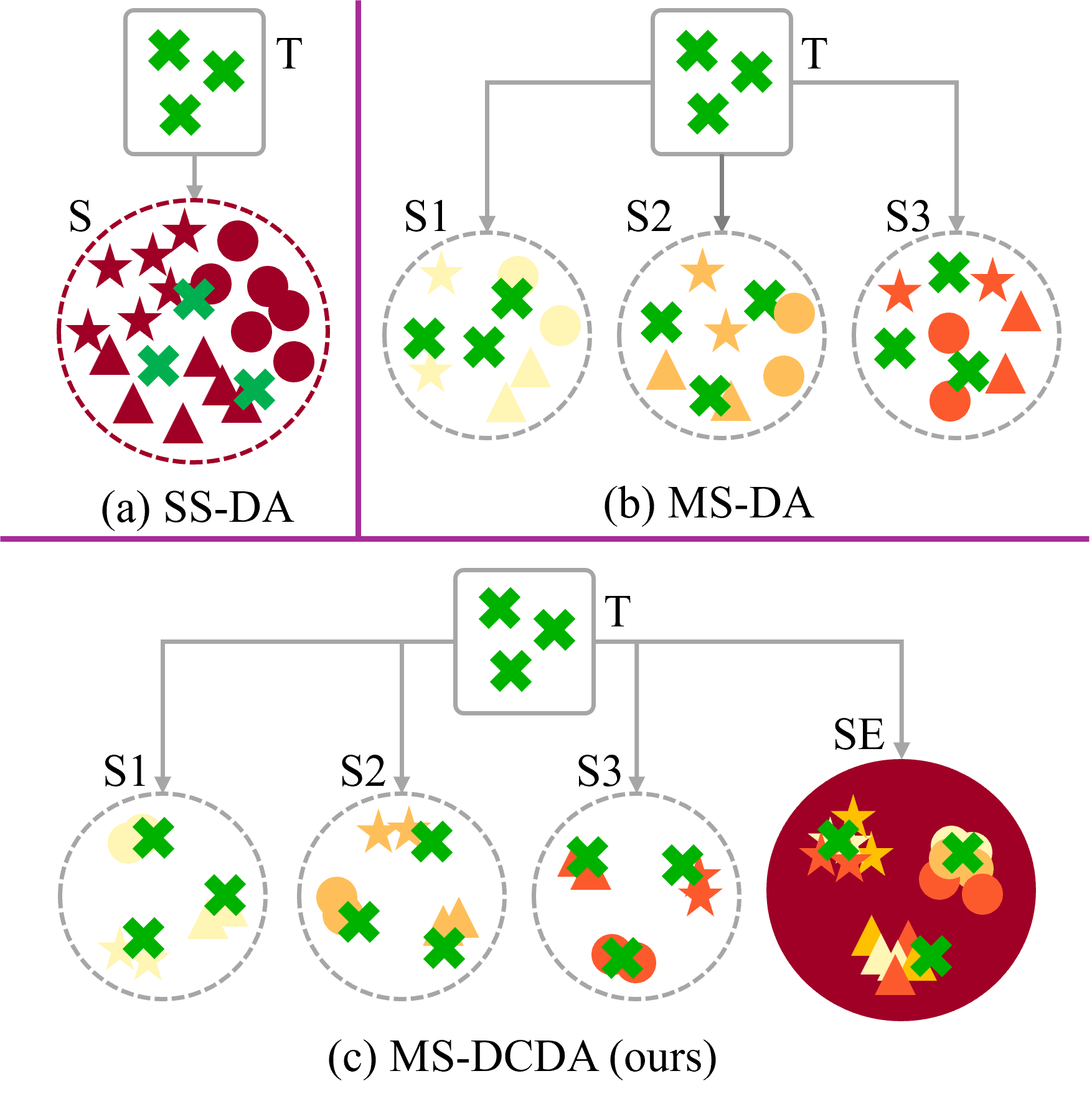}
\caption{Comparison of traditional discrepancy-based domain adaptation methods with our model. (a) Single-source domain adaptation (SS-DA) treats data from different subjects as a single source (S) and aligns target (T) with S, ignoring non-stationarity among individual sources. (b) Multi-source domain adaptation (MS-DA) aligns T with each individual source, but tends to produce sub-optimal results due to the lack of fine-grained alignment and imbalanced domain transferability and discriminability. (c) Our multi-source dynamic contrastive domain adaptation (MS-DCDA) model adapts T to individual sources and a complementary multi-source ensemble (SE) with class-awareness. Dynamically weighted domain transferability and discriminability provides improved accuracies, wider inter-class margin, and higher intra-class compactness.}

\label{fig: method_illus}
\end{figure}

Domain adaptation transfers knowledge from the source domain to the target domain, thereby minimizing the need for extensive data labeling \citep{zhu2021multi, zhu2022multisource, pan2009survey}. Unlike single-source domain adaptation (SS-DA), which treats EEG data from different subjects as a single source (see Fig. \ref{fig: method_illus}(a)), multi-source domain adaptation (MS-DA) considers each subject as an individual source domain (see Fig. \ref{fig: method_illus}(b)). The assumption of diverse data in the MS-DA may reduce domain bias and model overfitting \citep{zhu2019aligning, jiang2023deep}. However, domain shifts occur not only between each source and target, but also among different sources, potentially complicating the learning. Two common MS-DA include discrepancy-based and adversarial-based methods \citep{zhao2020multi}. The adversarial discriminative methods align target and source features through domain discriminator and adversarial objectives, whereas the discrepancy-based methods align features in latent space explicitly through discrepancy measures. Existing MS-DA approaches tend to assume shared features across domains, and their domain discriminability is limited to such coarse-grained global alignment. In addition, the typical static loss-weighting tends to produce suboptimal balance between the domain alignment and domain discriminability. 

 To address these challenges, we present a multi-source dynamic contrastive domain adaptation (MS-DCDA) method for EEG-based emotion recognition. Our model learns domain knowledge from both the source and target, where we consider not only the data contribution of each individual source but also a complementary source ensemble (see Fig. \ref{fig: method_illus}(c)). Additionally, our model also learns domain-variant features using fine-grained class-aware contrastive sub-domain adaptation. What is more, an optimal domain transferability and domain discrimination are achieved through a dynamically weighted loss function. Extensive evaluation of the proposed MS-DCDA is conducted on the SEED and SEED-IV dataset, and state-of-the-art performances are achieved for the cross-subject and the cross-session recognition respectively. In addition, the generalization performance of our model is evaluated through a dataset transfer study where key brain lobes involved in EEG emotion recognition are also examined. The specific contributions of our work are summarized as follows:
 
\begin{itemize}
\item We introduce a multi-source dynamic contrastive domain adaptation method, which models coarse-grained inter-domain and fine-grained intra-class adaptations through a multi-branch contrastive network and class-aware contrastive sub-domain discrepancy learning.

\item Our model leverages domain knowledge from each individual source and a complementary source ensemble uses dynamically weighted learning to achieve optimal domain transferability and discriminability, significant improvements in inter-class margin and intra-class compactness are demonstrated.

\item State-of-the-art performance is achieved on SEED and SEED-IV datasets, and our model outperforms the second best approach respectively by $1.5\%$ and $2.4\%$ in the cross-subject experiments as well as $2.2\%$ and $3.6\%$ in the cross-session experiment. 
\end{itemize}

\section{RELATED WORK}\label{sec2}
\subsection{EEG-based emotion recognition}
EEG provides a non-invasive measurement of the brain's electrophysiological activities. EEG signals mainly come from cortical pyramidal neurons perpendicular to the surface of the brain in the cerebral cortex. These signals offer numerous valuable clinical indications, including those related to higher cognitive functions such as emotions \citep{britton2016electroencephalography}. Two types of commonly used features for emotion classification include the time-domain features (e.g., standard deviation, mean, variance, and differential entropy) and the frequency-domain features (e.g., spectral power and power spectral density). In both feature domains, the use of differential entropy has demonstrated higher accuracy and greater stability in emotion classification than the power spectral density \citep{atkinson2016improving}. Traditional machine learning methods such as support vector machine \citep{weng2009computer, peng2009saliency, peng2017randomized, xiao2021dynamic}, K-nearest neighbor \citep{li2018emotion}, and random forest \citep{cheng2020emotion} make use of these manually crafted features for recognition. In recent years, machine learning and deep learning has achieved remarkable advancement in many areas \citep{sharifani2023machine, peng2024learning, peng2021cnn, peng2022computational, wang2024contextdet}, including affection computing and emotion recognition \citep{jafari2023emotion}. Deep learning methods learn features from data without manual feature extractions and selections. For example, convolutional neural networks {\citep{zhu_wang_zhang_2023}} have proven effective for lesion classification. CNNs have also been coupled with local information for emotion classification. By exploring frequency bands and channels via deep belief networks {\citep{zheng2015investigating}}, the accuracy has been improved on the SEED dataset. Additionally, using attention-based convolutional recurrent neural networks {\citep{tao}} for discriminative features extraction further boosting the performance of EEG based emotion recognition. Pre-trained Transformers are also optimized through random feature projection and head augmentation {\citep{zhu_liu_free_anjum_panneerselvam_2024}}, demonstrating improved efficiency, accuracy, and generalization in classification tasks.


\vspace{-0.5cm}
\subsection{Single-source domain adaptation}
Domain adaptation is a type of transductive transfer learning {\cite{pan2009survey}}, which adapts a model learned on source domain to classify data in target domain. Depending on the availability of class labels of samples in target domain during the training, deep visual domain adaptation are typically categorized into: supervised, semi-supervised, and unsupervised approaches. The case where the label is unavailable in target domain is considered as unsupervised domain adaptation (UDA) {\citep{wang2018deep}} {\cite{xu2022video}}{\citep{jimenez2023learning}}. Unsupervised domain adaptation has proven effective in addressing individual differences in EEG signals for emotion recognition, even in cases where the distribution of source and target domains are entirely different. To transfer cross-subject knowledge, maximum mean discrepancy (MMD) {\cite{gretton2006kernel}} has been widely adopted as a metric to describe the discrepancies between the distributions of source and target domains in UDA. For example, MMD preserves the domain invariance in deep adaptation network {\citep{li2018cross}} while addressing the individual differences. However, the naive MMD is not capable of addressing the class prior distributions, resulting in degraded domain adaptation accuracies. Class-aware subdomain adaptation network have been introduced \citep{kang2020contrastive, yan2017mind} to address the class weight bias. In addition, contrastive domain discrepancy metric \citep{li2022dynamic} is introduced for cross-subject and cross-session emotion recognition. The approximated joint marginal and conditional distributions \citep{li2019domain} also shows improved clustering and recognition accuracy on the SEED dataset. A deep subdomain adaptation network \citep{meng2022deep} captures fine-grained information from each class, improving the alignment of subdomains.


\subsection{Multi-source domain adaptation}
Single-source DA methods are limited to single-source domain distributions and relatively simple datasets. Multi- source domain adaptation algorithms are introduced to address more complicated data distributions. Unified convergent learning bound has been investigated for multi-source data \citep{ben2010theory}. The deep multi-source adaptation transfer network \citep{wang2021deep} combines an adaptation network  \citep{li2018cross} with a discriminator, allowing nonuniform distribution in cross-subject emotion recognition. The multi-source marginal distribution adaptation method \citep{atkinson2016improving} maps domains to a shared feature space using multi-layer perceptrons to extract unique domain-invariant features from each individual domain. The multi-kernel and multi-source MMD method \citep{gretton2012optimal} extends the single-source MMD to measure the differences between domains. Joint distribution was also investigated in multi-source domain adaptation \citep{she2023multisource} for fine-grained domain adaptation using reinforcement learning. Dynamic transfer learning \cite{li2021dynamic} is also explored to reduce multi-source to single-source adaptation thorough dynamic routing. 

\begin{figure*}[!t]
\centering
\includegraphics[width=7in]{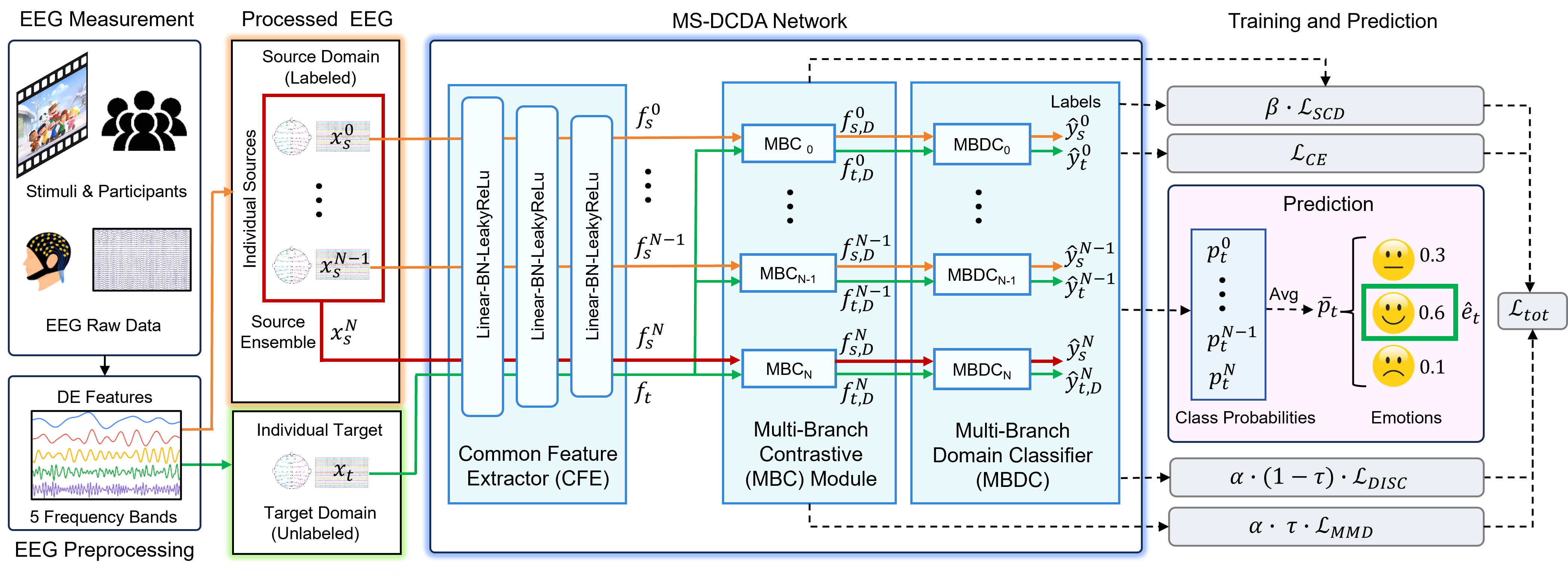}
\caption{The pipeline of EEG-based emotion recognition and the schematics of the proposed MS-DCDA model. The multi-source EEG data are measured and preprocessed. The five-band differential entropy (DE) features are extracted from the data of each participant. The multi-source dynamic contrastive domain adaptation (MS-DCDA)  model consists of three modules: the common feature extractor (CFE), the multi-branch contrastive (MBC) module, and the multi-branch domain classifier (MBDC). The CFE module extracts features extract domain-invariant features from source data $\{x_s^i| i=0,...N\}$ and the target data $x_t$ using a shared MLP. The MBC and MBDC modules are respectively comprised of $N+1$ independent branches $\{MBC_i\}$ and $\{MBDC_i\}$. The MBC module extracts the domain-variant features and the DC module classify them into domain-specific labels. During training, the $\mathcal{L}_{MMD}$ and $\mathcal{L}_{SCD}$ losses encourage class-independent and class-aware alignment respectively. The classification loss is comprised of the cross entropy loss $\mathcal{L}_{CE}$ and a complementary $\mathcal{L}_{DISC}$ loss, which encourages the predictions consistency across classifiers. Additionally, a dynamic coefficient $\tau$ optimizes the domain transferability and discriminability. During the test, the predicted target class probabilities are averaged across the classifiers, and the target emotion $\hat{e}_t$ is determined by the maximum mean class probability. We train our model in an unsupervised setting, where the source domain is labeled whereas the labels are unavailable for samples in the target domain.}
\label{model}
\vspace{-1em}
\label{fig: pipeline}
\end{figure*}

\section{METHODS}\label{sec3}
We introduce a multi-source dynamic contrastive domain adaptation (MS-DCDA) method for cross-subject and cross-session EEG-based emotion recognition. Our framework makes use of multi-branch contrastive neural net for class-aware domain alignment, which allows discriminative adjustment of marginal and conditional distributions of domains. The multi-source MMD is combined with a class-aware subdomain contrastive discrepancy (SCD) metric to tradeoff the coarse- and fine-grained domain alignments between source and target domains. Additionally, domain transferability and discriminability are optimally balanced through dynamically weighted domain alignment and classification penalties. The pipeline of EEG-based emotion recognition and the schematic of our MS-DCDA model are illustrated in Fig. {\ref{fig: pipeline}.  We learn transferable EEG features in an unsupervised domain adaptation (UDA) setting, where the samples are labeled in the source domain but remain unlabeled in the target domain during the training.}

\subsection{Data Preprocessing}
The pipeline of the emotion recognition starts with EEG data preprocessing, where the measured EEG signals are segmented and prefiltered. Differential entropy (DE) \citep{shi2013differential} features, which are effective in distinguishing low- and high-frequencies in EEG signals, are extracted from the EEG segments of each participant for five frequency bands (see Section \ref{sec:dataset}). For an EEG segment having a Gaussian distributed frequency spectrum $\mathcal{N}(\mu, \sigma^{2})$, its DE is given by:
\vspace{-1em}
\begin{equation}
\label{deqn_ex1}
\begin{aligned}
DE&=-\int_{X} f(x)log(f(x))dx \\
&=-\int_{-\infty}^{+\infty}{\frac{1}{ \sqrt{2\pi\sigma^{2}}}}e^{\frac{(x-\mu)^{2}}{2\sigma^{2}}}log({\frac{1}{ \sqrt{2\pi\sigma^{2}}}}e^{\frac{(x-\mu)^{2}}{2\sigma^{2}}})dx \\
&=\frac{1}{2}log(2\pi e\sigma^{2})
\end{aligned}
\end{equation}

\subsection{MS-DCDA Architecture}
The proposed MS-DCDA model consists of three modules: the common feature extractor, the multi-branch contrastive module, and the multi-branch domain classifier. The details of these modules are introduced below. 

{\bf{Common Feature Extractor}}. The common feature extractor (CFE) module extracts domain-invariant features from the source and target domain using a shared neural network. Here we use the precomputed DE features of EEG signals as pre-processed input. The target domain input is denoted as $X_t = \{x_t\}$, where $x_t\in\mathbb{R}^{W\times B}$, $W$ is number of EEG samples per individual subject, $B$ represents the total number of channels, which is the superposition of EEG channels in each frequency band. The source domain consists of $N+1$ sources $X_s = \{x_s^i|i = 0,...N\}$. The first $N$ source elements represent the individual sources $x_s^i\in\mathbb{R}^{W\times B}$ ($i<N$). The last source element indicates an ensemble of the individual sources $x_s^N = \bigcup_{i=0}^{N-1}\{x_s^i\}$. $x_s^N\in\mathbb{R}^{W_2 \times B}$ where $W_2$ is the total number of samples in the combined domain, which is the total number of samples in the first n-1 source domains. The common target feature is given by $F_t = \{f_t|f_t = CFE(x_t)\}$. The common source features are extracted not only from each individual source but also from the source ensemble $F_s= \{f_{s}^i\vert f_{s}^i = CFE(x_s^i), i = 0, \ldots ,N\}$. Here we employ multi-layer perceptron (MLP) for the CFE module, where the structure of each of the three layers is given respectively by {\small\fontfamily{pcr}\selectfont Linear310/256/128-BatchNorm1D-LeakyReLU(0.01)}.

{\bf{Multi-Branch Contrastive Module.}} The common features then pass through the multi-branch contrastive (MBC) module, which consists of $N+1$ branches $\{MBC_i| i = 0,...N\}$, each comprised of a single fully connected layer {\small\fontfamily{pcr}\selectfont Linear64-BatchNorm1D-LeakyReLU(0.01)}. Each MBC block projects a pair of a common source feature $F_s$ and a common target feature $F_t$ to their respective domain feature spaces, resulting a total of $2(N+1)$ domain features, where  $F_{s,D} = \{f_{s,D}^i| f_{s,D}^i = MBC_i(f_s^i), i = 0,...,N\}$ and $F_{t,D} = \{f_{t,D}^i| f_{t,D}^i = MBC_i(f_t), i = 0,...,N\}$.  \vspace{0.1cm}

{\bf{Multi-Branch Domain Classifier.}} The multi-branch domain classifier (MBDC) module predicts domain-specific labels from the features extracted by the MBC module. The MBDC module is comprised of $N+1$ linear classifiers $\{MBDC_i\}$ with structure {\small\fontfamily{pcr}\selectfont Linear32-softmax}. The predicted labels for each individual source and the ensemble source domain are denoted as $\widehat{Y}_s = \{\widehat{y}_s^i| \widehat{y}_s^i= MBDC_i(f_{s,D}^i), i = 0,...N\}$. The predicted label of the target is represented as $\widehat{Y}_t = \{\widehat{y}_t^i| \widehat{y}_t^i= MBDC_i(f_{t,D}^i), i = 0,...N\}$. The ground truth label is given by $Y_s=\{y_s^i|i=0,...,N\}$. During the training, $\hat{Y}_s$, $\hat{Y}_t$, and $Y_s$ are used in learning the domain classifier and serving as class guidances in learning fine-grained domain adaptation. At the prediction (or testing) stage, the emotion state of the target subject is identified from a set of $M$ defined emotions $\{e_m| m = 1,...,M\}$, each representing an emotion state (e.g., positive, negative, or neutral). The class probabilities of the target domain class $\{p_{t}^i = P(\hat{y_i}|y_i)| i = 0,...,N\}$ are averaged across the domain classifiers \citep{zhu2019aligning}, producing the mean class probabilities $\bar{p}_t = (\Sigma_{i=0}^{N} p_{t}^i)/N$, where $\bar{p}_t\in\mathbb{R}^{1\times M}$. The emotion state of the target participant is identified by the maximum mean class probability $\hat{e}_t = e_j$ where $j = \arg\max\limits_{m}\{\bar{p}_{t,m}\}$.

\subsection{MS-DCDA Learning}
The learning of the proposed MS-DCDA model involves four aspects: 1) coarse-grained domain aligment; 2) fine-grained sub-domain domain alignment with class awareness; and 3) fine-grained classification. 4) Dynamically weighted loss function for optimally balanced domain transferability and discrimination performance. 

\textbf{Domain Alignment}. The coarse-grained alignment of the source and target involves using the multi-source MMD to measure the difference between the MBC features of the pair-wise source and the target $\{f_{s,D}^i, f_{t,D}^i\}$. Mathematically, the MMD loss measures the difference between two distributions using their mean embeddings in the reproducing kernel Hilbert space. In practice, the squared value of MMD is typically estimated by kernel embedding:
\begin{equation}
\label{mmd}
\begin{aligned}
\mathcal{L}_{MMD}&\ = 
\frac{1}{N}\sum\limits_{i=0}^{N}\bigg(\frac{1}{N_{s}^2} \sum\limits_{u=1}^{N_{s}}\sum\limits_{v=1}^{N_{s}}k(f_{s,D}^{i,u},f_{s,D}^{i,v}) \\
& +\frac{1}{N_{t}^2} \sum\limits_{u=1}^{N_{t}}\sum\limits_{v=1}^{N_{t}}k(f_{t,D}^{i,u},f_{t,D}^{i,v}) \\
& -\frac{2}{N_{s}N_{t}} \sum\limits_{u=1}^{N_{s}}\sum\limits_{v=1}^{N_{t}}k(f_{s,D}^{i,u},f_{t,D}^{i,v})\bigg)\\
\end{aligned}
\end{equation}

\noindent where  $N_{s}$ and $N_{t}$ denote the number of mini-batch features sampled respectively from features of individual source and the target, $N$ represent the total number of sources-target feature pairs and $k$ is the Gaussian kernel function:
\begin{equation}
\begin{aligned}
\label{gaussian}
k(x,x') =\exp\bigg(-\frac{\lVert x-x' \rVert^2}{2\sigma^2}\bigg) 
\end{aligned}
\end{equation}

\textbf{Contrastive Sub-Domain Alignment.} To improve the inter-class margin and improve the intra-class compactness of the alignment, we extend the contrastive domain discrepancy \citep{kang2020contrastive} to a sub-domain contrastive discrepancy (SCD) for the alignment of our multi-branch contrastive features. The SCD measures feature differences of pair-wise MBC features $\{f_{s,D}^i, f_{t,D}^i\}$ in class level, which not only aligns inter-domain features in general but also aligns the intra-domain features with class-level accuracies:
\begin{equation}
\label{SCD}
\mathcal{L}_{SCD}=
\frac {1}{N(M+1)}\sum\limits_{i=0}^{N}\sum\limits_{c=0}^{M-1}\bigg(D_{cc}^i-\frac {1}{M}\sum\limits_{\substack{c'=0 \\ c'\neq c}}^{M-1}D_{cc'}^i\bigg)
\end{equation}

\noindent where M represents the number of emotional categories. Minimizing this loss encourages minimizing the intra-class discrepancies while maximizing the inter-class differences. For two classes c1 and c2 in a particular domain, the distance between the two classes is written as:
\begin{equation}
\label{D}
D_{c1c2}^i=d_{1}^i+d_{2}^i-2d_{3}^i
\end{equation}

\noindent Here MMD is employed to measure differences within the same subdomain. The MMD is not limited by data distribution types such as edge distribution or conditional distribution, making it favorable for class-level optimization:
\begin{equation}
\label{d1}
d_{1}^i=\sum\limits_{u=1}^{N_{s}}\sum\limits_{v=1}^{N_{s}}
\frac {g_{c1c1}(y_s^{i,u},y_s^{i,v})k(f_{s,D}^{i,u},f_{s,D}^{i,v})}{\sum_{u=1}^{N_{s}}\sum_{v=1}^{N_{s}}g_{c1c1}(y_s^{i,u},y_s^{i,v})} \hfill 
\end{equation}

\begin{equation}
\label{d2}
d_{2}^i=\sum\limits_{u=1}^{N_{t}}\sum\limits_{v=1}^{N_{t}}
\frac {g_{c2c2}(\widehat{y}_{t}^{i,u},\widehat{y}_{t}^{i,v})k(f_{t,D}^{i,u},f_{t,D}^{i,v})}{\sum_{u=1}^{N_{t}}\sum_{v=1}^{N_{t}}g_{c2c2}(\widehat{y}_{t}^{i,u},\widehat{y}_{t}^{i,v})}
\end{equation}

\begin{equation}
\label{d3}
d_{3}^i=\sum\limits_{u=1}^{N_{s}}\sum\limits_{v=1}^{N_{t}}
\frac {g_{c1c2}(y_{s}^{i,u},\widehat{y}_{t}^{i,v})k(f_{s,D}^{i,u},f_{t,D}^{i,v}))}{\sum_{u=1}^{N_{s}}\sum_{v=1}^{N_{t}}g_{c1c2}(y_{s}^{i,u},\widehat{y}_{t}^{i,v})}
\end{equation}

\noindent where $k(\cdot,\cdot)$ is the Gaussian kernel function (see Eq.\ref{gaussian}). The class masking function is given by:
\begin{equation}
{g_{c1c2}(y,y')}=\begin{cases}
1,&{\text{if}}\  {y = c1 \;\textrm{and}\;  y' = c2} \\ 
{0,}&{\text{otherwise.}} 
\end{cases}
\end{equation}

\noindent In cases where c1 = c2, the above equation measures the discrepancies within the same emotion class; for cases where c1 $\neq$ c2, it measures the discrepancies between the two classes. Penalizing the SCD loss minimizes the intra-class discrepancies while maximizing the inter-class discrepancies. Due to the lack of target label, here we use pseudo labels $\widehat{y}_{t}^{i}$ as alternatives to estimate the SCD. Although the approximated pseudo labels might not be perfect, the impact of label noise on the SCD loss is typically minimal compared to a large dataset \citep{kang2020contrastive}. The model robustness to pseudo label errors may attribute to the the mean embedding of the distribution within a reproducing kernel Hilbert space of MMD.\vspace{0.1cm} 

\textbf{Classification and Feature Discrimination.}
Given the pair-wise multi-branch features $\{f_{s,D}^i, f_{t,D}^i\}$, we train the classifier using cross entropy loss:
\begin{equation}
\label{Ce}
\mathcal{L}_{CE}=-\frac {1}{N}\sum\limits_{i=0}^{N}y_{s}^{i}\log{P(\widehat{y}_{s}^{i}|x_s^i)}
\end{equation}
where $P($·$|$·$)$ represents the probability distribution of the predicted label. Using the cross entropy loss alone typically leads to imbalanced prediction across different classifiers. To encourage consistent prediction of the target and accelerate the convergence rate of our multi-branch domain classifiers, a complementary classification loss $L_{DISC}$ is introduced:
\begin{equation}
\label{disc}
\mathcal{L}_{DISC}=
\frac {1}{N} \bigg(\sum\limits_{i\neq j}^{N}E_{x\sim x_{t}}\mid \widehat{y}_t^i-\widehat{y}_t^j \mid \bigg)
\end{equation}
where $E_{x\sim x_{t}}$ represents randomly selection of samples from the target features. \vspace{0.1cm}


\begin{algorithm}[t]
\caption{MS-DCDA for class-aware cross-subject, and cross-session emotion recognition from EEG signals.}
\label{algorithm}
\begin{algorithmic}
\STATE 
\STATE {\textbf{Input: Pre-processsed EEG signals}}
\STATE {$X_s=\{ x_s^i|i=1,...,N \}$ : Source domain data.\\
$Y_s=\{ y_s^i|i=0,...,N \}$: Source labels. \\
$x^{t}$: Unlabeled target domain data.\\
$\{e_m| m = 1,...M\}$: Emotion categories\\
$N_s, N_t$: Batch size of sampling source and target features\\
$Q$: Total number of iterations\\
$M$: Total number of emotion categories}
\STATE {\textbf{Output: The recognized emotion category $\hat{e}_t$}}
\STATE \textbf{for} $iter$ = 0, . . . , $Q-1$ \textbf{do} 
\STATE \hspace{0.3cm} Extract target common feature $f_{t} = CFE(x_t)$  
\STATE \hspace{0.3cm} \textbf{for} $i$ = 0, . . . , $N$ \textbf{do} 
\STATE \hspace{0.6cm} Extract source common features $f^{i}_{s} = CFE(x_s^i)$
\STATE \hspace{0.6cm} Extract source domain features $f^{i}_{s,D} = MBC_i(f_s^i)$ 
\STATE \hspace{0.6cm} Extract target domain features $f^{i}_{t,D} = MBC_i(f_t)$ 
\STATE \hspace{0.6cm} Classify source domain labels $\hat{y}^{i}_{s} = MBDC_i(f^{i}_{s,D})$ 
\STATE \hspace{0.6cm} Classify target domain labels $\hat{y}^{i}_{t} = MBDC_i(f^{i}_{t,D})$ 
\STATE \hspace{0.3cm} \textbf{end} 
\STATE \hspace{0.3cm} Sample batch $\{f^{i,u}_{s,D}|i = 0,...N, u = 1,...N_s\}$ 
\STATE \hspace{0.3cm} Sample batch $\{f^{i,v}_{t,D}|i = 0,...N, v = 1,...N_t\}$ 
\STATE \hspace{0.3cm} Update loss $\mathcal{L}_{MMD}$ using Eq.\ref{mmd}
\STATE \hspace{0.3cm} Sample batch $\{\hat{y}^{i,u}_{s}|i = 0,...N, u = 1,...N_s\}$ 
\STATE \hspace{0.3cm} Sample batch $\{\hat{y}^{i,v}_{t}|i = 0,...N, v = 1,...N_t\}$ 
\STATE \hspace{0.3cm} Update loss $\mathcal{L}_{SCD}$, $\mathcal{L}_{CE}$, $\mathcal{L}_{DISC}$ using Eq.\ref{SCD}-\ref{disc}
\STATE \hspace{0.3cm} Update dynamic weight $\tau$ using Eq.\ref{dynamic balance factor}
\STATE \hspace{0.3cm} Update total loss $\mathcal{L}_{tot}$ using Eq.\ref{LOSS}
\STATE \hspace{0.3cm} Backpropagate, update $CFE$, $\{MBC_i\}$, $\{MBDC_i\}$ 
\STATE \textbf{end} 
\STATE Obtain recognition probab. $\{p_{t}^i = P(\hat{y_i}|y_i)| i = 0,...,N\}$
\STATE Compute the average probability $\bar{p}_t = (\Sigma_{i=0}^{N} p_{t}^i)/N$ 
\STATE Identify emotion $\hat{e}_t = e_j$ for $\arg\max\limits_{j}\{\bar{p}_{t,m}|m = 1,...,M\}$
\end{algorithmic}
\end{algorithm}

\textbf{Dynamic Weighted Learning.}
Learning domain discriminability is typically prioritized for data that features small inter-domain variations. Conversely, improving domain transferability becomes more in favor for data with larger inter-domain variations. To achieve an optimal tradeoff between the domain transferability and discriminability, we adopt a dynamic weighted learning method \citep{xiao2021dynamic}, where a dynamic coefficient is used to adjust the losses of domain feature alignment and domain feature discrimination. In the dynamic learning, the multi-source MMD (see Eq.\ref{mmd}) is employed as an indicator of the cross-domain alignment. Using the linear discriminant analysis, domain discriminability is measured from features extracted by the CFE module:
\begin{equation}
\label{LDA}
\arg\max\limits _{W} J(W)=\frac{Tr(W^{T}S_{b}W)}{Tr(W^{T}S_{W}W)}
\end{equation}

\noindent where $S_{b}$ and $S_{W}$ are respectively the inter and intra class variance of the CFE features, and $Tr(\cdot)$ represents trace of the two variance matrices \citep{chen2019transferability}. A larger $J(W)$ value indicates a better discriminability, while a smaller $L_{MMD}$ value indicates better a domain transferability. The dynamic coefficient $\tau$ which balances the domain discriminability and transferability is given by:  

\begin{equation}
\label{dynamic balance factor}
\tau=\frac{L_{MMD}}{L_{MMD}+1-{J}(W)}
\end{equation}

\noindent where the values of $L_{MMD}$ and $J(W)$ are normalized to the range of [0,1] with their respective maximum and minimum. The dynamic coefficient $\tau$ varies between 0 and 1. A large $\tau$ indicates that the model favors domain alignment, whereas a smaller $\tau$ reflects a preference for class discriminability.

The training objective for MS-DCDA consists of the domain alignment learning, class-level domain alignment learning, class discrimination learning, and classifier learning, where the total loss is dynamically weighted as follows:
\begin{equation}
\label{LOSS}
\mathcal{L}_{tot}=\mathcal{L}_{CE}+\alpha\cdot\big((1-\tau)L_{DISC}+\tau \mathcal{L}_{MMD}\big) + \beta\cdot\mathcal{L}_{SCD}
\end{equation}

\noindent In addition to the dynamic coefficient $\tau$, another two dynamically changing coefficients include $\alpha$=($\frac{2}{1+e^{-10{\times}iter/epoch}}$)-1 and $\beta$=$\frac{\alpha}{10}$, where $iter$ is the index of iteration of training. The values of $\alpha$ and $\beta$ increase from 0 to 1 at different speeds as the number of iterations increases.  The MS-DCDA algorithm is summarized in Algorithm \ref{algorithm}.

\section{EXPERIMENTS}\label{sec4}
This section introduces the datasets, experimental setup, and training details of the proposed MS-DCDA model. 


\subsection{Dataset}
\label{sec:dataset}
We evaluate baseline methods and our MS-DCDA model on the Shanghai Jiao Tong University (SJTU) Emotion EEG Dataset (SEED) \citep{zheng2015investigating}\citep{duan2013differential} and the SJTU Emotion EEG Dataset IV (SEED-IV) \citep{zheng2018emotionmeter}. Fifteen healthy participants (subjects), including seven males and  eight females with an average age of 23 years, participated in data collection. The {\bf{SEED}} dataset is collected by having participants watch 15 movie clips, which incite positive, neutral, and negative emotions. While subjects were watching movie clips, EEG data were obtained through a 62-channel ESI NeuroScan system. The collected data was downsampled to 200 Hz. The SEED signals are filtered with bandpass filter of 0-75Hz and segmented into non-overlapping intervals of 1 second each. The {\bf{SEED-IV}} dataset is collected by having the same participants watch 24 movie clips, which incite neutral, sad, fear, and happy emotions. The SEED-IV signals are prefiltered by a 1-75 Hz bandpass filter and segmented into non-overlapping intervals of 4 seconds each. For both the SEED and SEED-IV dataset, signals are collected in three sessions for each subject and each segment.  

{\bf{Data Preprocessing. }} Each of the filtered EEG segments is considered as a data sample in training. DE features are extracted from each segment for five frequency bands: delta (1-4 Hz), theta (4-8 Hz), alpha (8-14 Hz), beta (14-31 Hz), and gamma (31-50 Hz), resulting in a feature dimension of 62 channels $\times$ 5 frequency bands. A total of 3394 EEG samples are collected for each participant per session for the SEED set. The SEED-IV set consists of 851, 832, 822 samples per participant for each of the three sessions respectively. Labels are generated for both datasets, and pre-processed EEG data of each participant are normalized electrode-wise \citep{chen2021ms} based on individual mean and standard deviation values.

\subsection{Experimental Setup}
\label{sec:exp_setup}
To evaluate the performance of the proposed MS-DCDA network, we performed three round leave-one-out (LOO) cross validations for cross-subject and cross-session experiments respectively. In the \textbf{Cross-Subject} experiment, a total number of 15 subjects is evaluated, where EEG of one subject is selected as the target and EEG of the remaining 14 subjects are considered as sources. EEG signals of subjects measured in one session are selected as target, and EEG signals of subjects measured from the rest two sessions are considered as source. 


\subsection{Training}
The proposed MS-DCDA model is trained with Adam optimizer for 50 epochs. We use a batch size of 32 and 16 respectively for the SEED and SEED-IV dataset. The learning rate remains at 5 $\times$ $10^{-3}$ for both. The dynamic coefficients are given by $\alpha$=($\frac{2}{1+e^{-10{\times}iter/epoch}}$)-1 and $\beta$=$\frac{\alpha}{10}$ respectively, where $iter$ is the index of training iterations. All the training tasks are conducted on a NVIDIA GeForce RTX 3060 GPU. The implementation is based on Python 3.9.18, PyTorch 2.0.0, and Torchvision 0.15.0.

\captionsetup{font=normalsize}
\begin{table*}[htbp]
\caption{Cross-subject Results of a Single Round of Leave-one-out (LOO) Validation}
\setlength{\tabcolsep}{2.7pt} 
\small
\centering
\begin{tabularx}{\textwidth}{@{}>{\hspace{100pt}}l*{16}{c}@{}} 
\toprule
        \multicolumn{17}{c}{SEED}\\
        \multicolumn{1}{c}{Method}  &S1     &      S2 &      S3 &      S4 &     S5 &
        S6 &      S7 &      S8 &      S9 &      S10  &
        S11 &      S12 &      S13 &      S14 &      S15&      Avg\\
        \midrule
        \multicolumn{1}{l}{DDC \citep{tzeng2014deep}}&72.78&88.57&62.61&\cellcolor{LightGray}\textbf{81.59}&77.02&99.56&81.76&72.33&87.57&82.82&73.90&82.76&81.14&94.52&78.93&81.19\\
        \multicolumn{1}{l}{DCORAL \citep{sun2016deep}}&81.65&76.78&79.32&75.63&79.88&81.70&73.10&79.55&74.93&82.47&80.26&76.84&85.21&75.37&75.63&78.56\\
        \multicolumn{1}{l}{DANN \citep{kang2020contrastive}}&71.77&86.53&61.67&81.26&78.29&99.77&77.34&72.01&93.02&77.28&73.40&83.50&81.85&87.10&83.03&80.52\\
        \multicolumn{1}{l}{DAN \citep{li2018cross}}&\cellcolor{LightGray}\textbf{84.74}&85.48&80.73&76.16&87.63&87.86&79.02&82.74&81.56&87.86&80.00&75.93&80.14&81.29&81.47&82.17\\
        \multicolumn{1}{l}{MS-MDA \citep{chen2021ms}}&74.37&97.67&84.77&77.52&84.92&95.85&\cellcolor{LightGray}\textbf{90.34}&74.51&90.78&79.73&82.65&89.28&84.33&97.00&91.34&86.34\\
      \multicolumn{1}{l}{ MS-ADA \citep{she2023multisource}}&87.74&91.22&\cellcolor{LightGray}\textbf{100.00}&77.34&\cellcolor{LightGray}\textbf{95.49}&80.08&82.15&79.41&86.01&\cellcolor{LightGray}\textbf{98.68}&83.44&86.01&89.45&79.52&89.51&87.07\\
        \multicolumn{1}{l}{\cellcolor{LightGray}\textbf{MS-DCDA (ours)}}&76.58&\cellcolor{LightGray}\textbf{98.70}&75.99&79.11&90.51&\cellcolor{LightGray}\textbf{100.00}&83.53&\cellcolor{LightGray}\textbf{88.21}&\cellcolor{LightGray}\textbf{100.00}&88.80&\cellcolor{LightGray}\textbf{95.58}&\cellcolor{LightGray}\textbf{99.29}&\cellcolor{LightGray}\textbf{92.55}&\cellcolor{LightGray}\textbf{94.87}&\cellcolor{LightGray}\textbf{99.00}&\cellcolor{LightGray}\textbf{90.85}\\
        \midrule
        \multicolumn{17}{c}{SEED-IV}\\
        \midrule
        \multicolumn{1}{l}{DDC \citep{tzeng2014deep}} &62.62&63.10&57.21&68.99&57.21&79.09&71.75&72.96&72.96&59.50&68.75&65.38&68.99&70.67&67.31&67.14\\
        \multicolumn{1}{l}{DCORAL \citep{sun2016deep}}&66.54&53.91&59.38&60.81&64.19&68.62&71.61&82.16&74.48&61.20&75.52&62.24&75.00&71.22&70.31&67.81\\
        \multicolumn{1}{l}{DANN \citep{kang2020contrastive}}&62.50&59.38&54.09&56.61&64.66&75.12&78.13&75.96&90.05&61.30&75.96&65.87&74.88&70.67&62.86&67.87\\
        \multicolumn{1}{l}{DAN \citep{li2018cross}}&61.66&54.81&62.26&65.02&57.45&65.02&71.88&64.78&76.20&58.29&66.83&\cellcolor{LightGray}\textbf{68.99}&68.27&69.35&59.62&64.70\\
        \multicolumn{1}{l}{MS-MDA \citep{chen2021ms}}&74.52&52.28&67.43&\cellcolor{LightGray}\textbf{74.04}&72.48&78.37&\cellcolor{LightGray}\textbf{91.47}&\cellcolor{LightGray}\textbf{91.11}&85.70&71.63&75.00&66.47&\cellcolor{LightGray}\textbf{91.47}&77.52&72.12&76.11\\
         \multicolumn{1}{l}{MS-ADA \citep{she2023multisource}}&68.15&60.82&54.33&56.97&73.80&96.27&83.41&69.47&\cellcolor{LightGray}\textbf{91.47}&58.29&\cellcolor{LightGray}\textbf{77.04}&48.80&53.49&75.72&70.43&69.23\\
        \multicolumn{1}{l}{\cellcolor{LightGray}\textbf{MS-DCDA (ours)}} &\cellcolor{LightGray}\textbf{82.93}&\cellcolor{LightGray}\textbf{71.27}&\cellcolor{LightGray}\textbf{77.16}&73.32&\cellcolor{LightGray}\textbf{79.45}&\cellcolor{LightGray}\textbf{95.55}&79.69&88.94&78.37&\cellcolor{LightGray}\textbf{71.75}&76.68&63.22&78.73&\cellcolor{LightGray}\textbf{83.41}&\cellcolor{LightGray}\textbf{76.92}&\cellcolor{LightGray}\textbf{78.49}\\
        \bottomrule
\end{tabularx}
\vspace{-0.5em}
\label{long_tab_cross_sub}
\end{table*}

\captionsetup{font=normalsize}
\begin{table}[htbp]
\caption{Cross-Subject Results of Three LOO Rounds}
\setlength{\tabcolsep}{4.8pt} 
\small
\raggedright  
\begin{tabular}{@{}>{\hspace{6pt}}l*{5}{c}@{}} 
\toprule
\multicolumn{5}{c}{SEED}\\
\multicolumn{1}{c}{Method}  & ACC-mean & ACC-best & F1 & Kappa\\
\midrule
DDC \citep{tzeng2014deep} &80.45&81.19$\pm$8.89&0.81&0.72\\
DCORAL \citep{sun2016deep} &77.04&78.56$\pm$7.78&0.80&0.70\\
DANN \citep{kang2020contrastive}&79.56&80.52$\pm$8.96&0.81&0.71\\
DAN \citep{li2018cross}&80.95&82.17$\pm$3.74&0.82&0.73\\
MS-MDA \citep{chen2021ms} &85.74&86.34$\pm$7.43&0.86&0.79\\
MS-ADA \citep{she2023multisource} &84.18&87.07$\pm$6.84&0.87&0.81\\
\cellcolor{LightGray}\textbf{MS-DCDA (ours)}&\cellcolor{LightGray}\textbf{88.75}&\cellcolor{LightGray}\textbf{90.84$\pm$8.31}&\cellcolor{LightGray}\textbf{0.91}&\cellcolor{LightGray}\textbf{0.86}\\
\midrule
\multicolumn{5}{c}{SEED-IV}\\
\midrule
DDC \citep{tzeng2014deep} &65.22&67.14$\pm$6.03&0.62&0.54\\
DCORAL \citep{sun2016deep} &65.72&67.81$\pm$7.29&0.65&0.57\\
DANN \citep{kang2020contrastive} &63.78&67.87$\pm$8.14&0.67&0.58\\
DAN \citep{li2018cross} &57.64&64.70$\pm$5.62&0.59&0.51\\
MS-MDA \citep{chen2021ms} &71.22&76.11$\pm$10.29&0.55&0.40\\
MS-ADA \citep{she2023multisource} &65.95&69.23$\pm$13.62&0.67&0.57\\
\cellcolor{LightGray}\textbf{MS-DCDA (ours)}&\cellcolor{LightGray}\textbf{74.34}&\cellcolor{LightGray}\textbf{78.49$\pm$7.36}&\cellcolor{LightGray}\textbf{0.79}&\cellcolor{LightGray}\textbf{0.71}\\
\bottomrule
\end{tabular}
\vspace{-1em}
\label{short_tab_cross_sub}
\end{table}

\captionsetup{font=normalsize}
\begin{table}[htbp]
\caption{Cross-Subject Results of Three LOO Rounds (ACC-best)}
\setlength{\tabcolsep}{14.2pt} 
\small
\raggedright  
\begin{tabular}{@{}>{\hspace{6pt}}l*{3}{c}@{}} 
\toprule
\multicolumn{1}{c}{Method}  & SEED & SEED-IV \\
\midrule
BiDANN \cite{li2018bi}&	84.14$\pm$6.87	& 65.59$\pm$10.39\\
RGNN \cite{zhong2020eeg}&    85.30$\pm$6.72&73.84$\pm$8.02 \\
TANN \cite{li2021novel}&	84.41$\pm$8.75	&68.00$\pm$8.35\\
PPDA \cite{zhao2021plug}&	86.70$\pm$7.10	&-\\
GMSS \cite{li2022gmss}&	86.52$\pm$6.22 	&73.48$\pm$7.41\\
HVF\textsubscript{2}N-DBR \cite{guo2022horizontal}&	89.33$\pm$10.13	&73.60$\pm$2.91\\
MWACN \cite{zhu2022multisource}&	89.30$\pm$9.18  	&74.60$\pm$10.77\\
PCDG \cite{cai2023two}&	87.30$\pm$2.10 	&73.60$\pm$5.10\\
MFA-LR \cite{jimenez2023learning}&	89.11$\pm$7.72	&74.99$\pm$12.1\\
\cellcolor{LightGray}\textbf{MS-DCDA (ours)} &\cellcolor{LightGray}\textbf{90.84$\pm$8.31}	&\cellcolor{LightGray}\textbf{78.49$\pm$7.36}\\
\bottomrule
\end{tabular}
\vspace{-1em}
\label{sota_tab_cross_sub}
\end{table}

\section{RESULTS}
Three rounds of LOO cross-validations (see Section \ref{sec:exp_setup}) are conducted for cross-subject and cross-session experiments respectively. The same validations are repeated for the SEED and SEED-IV datasets respectively. For each dataset and experiment, we evaluated the mean recognition accuracy (ACC-mean) and the best recognition accuracy (ACC-best) across the three LOO validation rounds for the target domain. The best recognition accuracies are given in the form of mean $\pm$ standard deviation for the three rounds of LOO. 

We compare the proposed MS-DCDA with several advanced methods for both cross-subject and cross-session experiments. They include four single-source domain adaptation methods: deep domain confusion (DDC) \citep{tzeng2014deep}, DAN \citep{li2018cross}, DANN \citep{kang2020contrastive}, deep correlation alignment (DCORAL) \citep{sun2016deep}; Two multi-source domain adaptation models: multisource marginal distribution adaptation (MS-MDA) \citep{chen2021ms}, and multi-source associate domain adaptation (MS-ADA) \citep{she2023multisource}. The DDC method minimizes the domain distribution discrepancies between the source and the target using MMD. The DCORAL model aligns the second-order statistics of domain distributions using linear transformations. The DANN aligns domain feature distributions through backpropagation. The DAN method aligns high-level domain features through the use of deep neural networks and multi-kernel MMD. In the MS-MDA method, multi-branch networks and MMD are combined for the multi-source alignment. The MS-ADA aligns the edge distributions and through MMD and reinforcement learning respectively. In addition, we compare our MS-DCDA against several more models for the cross-subject experiment. They include BiDANN \citep{li2018bi}, RGNN \citep{zhong2020eeg}, TANN \citep{li2021novel}, PPDA \citep{zhao2021plug}, GMSS \citep{li2022gmss}, HVF\textsubscript{2}N-DBR \citep{guo2022horizontal}, MWACN \citep{zhu2022multisource}, PCDG \citep{cai2023two}, and MFA-LR \citep{jimenez2023learning}.


\captionsetup{font=normalsize}
\begin{table*}[htbp]
\caption{Cross-session Results of a Single Round of Leave-one-out (LOO) Validation}
\setlength{\tabcolsep}{1.65pt} 
\small
\centering
\begin{tabularx}{\textwidth}{@{}>{\hspace{100pt}}l*{16}{c}@{}} 
\toprule
        \multicolumn{17}{c}{SEED}\\
        \multicolumn{1}{c}{Method}  &S1     &      S2 &      S3 &      S4 &     S5 &  S6 &      S7 &      S8 &      S9 &      S10  &
        S11 &      S12 &      S13 &      S14 &      S15&      Avg\\
        \midrule
        \multicolumn{1}{l}{DDC \citep{tzeng2014deep}}&93.64&94.99&91.04&96.67&93.70&98.85&78.82&90.19&83.32&81.67&76.78&92.55&90.51&97.17&87.13&89.80\\
        \multicolumn{1}{l}{DCORAL \citep{sun2016deep}}&82.21&93.03&90.90&87.26&84.19&97.66&83.05&89.63&80.32&85.58&77.82&90.02&90.90&97.51&79.39&87.30\\
        \multicolumn{1}{l}{DANN \citep{kang2020contrastive}}&90.36&96.14&91.42&96.02&95.28&97.97&79.13&90.74&81.90&84.94&78.01&89.33&90.45&99.00&80.84&89.43\\
        \multicolumn{1}{l}{DAN \citep{li2018cross}}&91.42&\cellcolor{LightGray}\textbf{100.00}&90.92&92.39&98.70&\cellcolor{LightGray}\textbf{100.00}&82.25&95.25&85.97&82.16&81.10&\cellcolor{LightGray}\textbf{100.00}&92.13&99.73&89.00&92.07\\
        \multicolumn{1}{l}{MS-MDA \citep{chen2021ms}}&\cellcolor{LightGray}\textbf{94.43}&90.78&91.34&\cellcolor{LightGray}\textbf{100.00}&\cellcolor{LightGray}\textbf{100.00}&\cellcolor{LightGray}\textbf{100.00}&84.12&97.17&92.78&92.13&84.03&91.13&96.61&97.61&91.28&93.61\\
        \multicolumn{1}{l}{MS-ADA \citep{she2023multisource}}&84.55&93.12&98.44&92.10&97.73&\cellcolor{LightGray}\textbf{100.00}&83.87&\cellcolor{LightGray}\textbf{100.00}&\cellcolor{LightGray}\textbf{98.11}&91.27&79.36&85.55&95.05&\cellcolor{LightGray}\textbf{100.00}&\cellcolor{LightGray}\textbf{100.00}&93.28\\
        \multicolumn{1}{l}{\cellcolor{LightGray}\textbf{MS-DCDA (ours)}}&88.51&90.75&\cellcolor{LightGray}\textbf{100.00}&90.95&\cellcolor{LightGray}\textbf{100.00}&93.96&\cellcolor{LightGray}\textbf{100.00}&99.73&84.68&\cellcolor{LightGray}\textbf{95.82}&\cellcolor{LightGray}\textbf{99.53}&95.76&\cellcolor{LightGray}\textbf{99.03}&\cellcolor{LightGray}\textbf{100.00}&98.59&\cellcolor{LightGray}\textbf{95.82}\\
        \midrule
        \multicolumn{17}{c}{SEED-IV}\\
        \midrule
        \multicolumn{1}{l}{DDC \citep{tzeng2014deep}} &74.75&59.88&66.13&61.88&92.50&90.25&\cellcolor{LightGray}\textbf{81.50}&81.75&57.75&84.25&\cellcolor{LightGray}\textbf{86.75}&91.88&91.50&83.50&66.00&78.02\\
        \multicolumn{1}{l}{DCORAL \citep{sun2016deep}}&74.74&66.41&64.71&73.44&95.44&82.81&71.22&82.29&54.04&85.68&76.69&86.33&89.58&79.82&56.90&76.01\\
        \multicolumn{1}{l}{DANN \citep{kang2020contrastive}}&73.32&51.32&68.27&76.20&87.02&94.23&66.47&80.89&83.17&78.61&69.47&79.57&90.02&73.68&57.57&75.32\\
        \multicolumn{1}{l}{DAN \citep{li2018cross}}&70.13&59.63&45.13&73.50&92.25&86.75&78.50&91.00&56.88&\cellcolor{LightGray}\textbf{82.88}&74.63&\cellcolor{LightGray}\textbf{93.00}&92.38&80.13&64.63&76.09\\
        \multicolumn{1}{l}{MS-MDA \citep{chen2021ms}}&75.36&51.80&73.92& \textbf{85.46}&78.00&93.15&71.51&87.02&\cellcolor{LightGray}\textbf{85.46}&78.00&81.97&72.36&85.82&86.78&72.36&78.60\\
         \multicolumn{1}{l}{MS-ADA \citep{she2023multisource}}&64.13&56.63&61.00&60.13&\cellcolor{LightGray}\textbf{98.00}&93.38&76.75&81.88&60.88&64.63&78.88&86.13&\cellcolor{LightGray}\textbf{100.00}&82.50&\cellcolor{LightGray}\textbf{77.63}&76.17\\
        \multicolumn{1}{l}{\cellcolor{LightGray}\textbf{MS-DCDA (ours)}} &\cellcolor{LightGray}\textbf{81.37}&\cellcolor{LightGray}\textbf{73.16}&\cellcolor{LightGray}\textbf{69.24}&66.79&97.92&\cellcolor{LightGray}\textbf{92.65}&72.43&\cellcolor{LightGray}\textbf{94.98}&68.01&79.53&84.68&90.56&95.22&\cellcolor{LightGray}\textbf{94.85}&72.30&\cellcolor{LightGray}\textbf{82.25}\\
        \bottomrule
\end{tabularx}
\vspace{-0.5em}
\label{long_tab_cross_session}
\end{table*}

\captionsetup{font=normalsize}
\begin{table}[htbp]
\caption{Cross-Session Results of Three Leave-one-out (LOO) Rounds}
\setlength{\tabcolsep}{4.8pt} 
\small
\raggedright  
\begin{tabular}{@{}>{\hspace{6pt}}l*{5}{c}@{}} 
\toprule
\multicolumn{5}{c}{SEED}\\
\multicolumn{1}{c}{Method}  & ACC-mean & ACC-best & F1 & Kappa\\
\midrule
    DDC \citep{tzeng2014deep}&89.15&89.80$\pm$6.61&0.88&0.81\\
    DCORAL \citep{sun2016deep}&85.57&87.30$\pm$6.00&0.87&0.81\\
    DANN \citep{kang2020contrastive}&88.14&89.43$\pm$6.74&0.88&0.81\\
    DAN \citep{li2018cross}&90.60&92.07$\pm$6.66&0.94&0.90\\
    MS-MDA \citep{chen2021ms}&91.80&93.61$\pm$4.95&0.94&0.91\\
    MS-ADA \citep{she2023multisource} &92.02&93.28$\pm$6.72&0.93&0.89\\
    \cellcolor{LightGray}\textbf{MS-DCDA(ours)}&\cellcolor{LightGray}\textbf{93.92}&\cellcolor{LightGray}\textbf{95.82$\pm$4.81}&\cellcolor{LightGray}\textbf{0.95}&\cellcolor{LightGray}\textbf{0.93}\\
\midrule
\multicolumn{5}{c}{SEED-IV}\\
\midrule
    DDC \citep{tzeng2014deep}&75.30&78.02$\pm$12.11&0.77&0.70\\
    DCORAL \citep{sun2016deep}&72.86&76.01$\pm$11.44&0.74&0.67\\
    DANN \citep{kang2020contrastive}&72.97&75.32$\pm$11.23&0.74&0.67\\
    DAN \citep{li2018cross}&72.79&76.09$\pm$14.12&0.74&0.68\\
    MS-MDA \citep{chen2021ms}&75.78&78.60$\pm$9.61&0.73&0.66\\
    MS-ADA \citep{she2023multisource}&72.22&76.17$\pm$13.91&0.77&0.69\\
\cellcolor{LightGray}\textbf{MS-DCDA (ours)}&\cellcolor{LightGray}\textbf{77.53}&\cellcolor{LightGray}\textbf{82.25$\pm$11.00}&\cellcolor{LightGray}\textbf{0.83}&\cellcolor{LightGray}\textbf{0.77}\\
\bottomrule
\end{tabular}
\vspace{-2em}
\label{short_tab_cross_session}
\end{table}
\subsection{Cross-Subject Results}
The comparisons of the emotion recognition results of our MS-DCDA model and the baseline algorithms are presented in Tables \ref{long_tab_cross_sub}, where the best results are highlighted in bold font. The alternative methods are tuned to their best performances. If the relevant parameters are published, we will use them. If not, the parameters will remain consistent with our method. The cross-subject experimental results on SEED and SEED-IV, which are based on both the state-of-the-art methods and our proposed method, are presented in Table \ref{short_tab_cross_sub}. On the SEED dataset, the constructed MS-DCDA model significantly improves recognition performance, achieving 90.84\% for the third session and 88.75\% for the average of three sessions. In comparison to other methods, such as MS-MDA and MS-ADA, both being multiple source domain adaptation methods, our method demonstrates superior performance. Compared to MS-MDA which achieves the highest average accuracy, our method has improved accuracy by 3.01\%. Compared to MS-ADA which achieves the highest average accuracy, our method  has improved accuracy by 3.77\%. On the SEED-IV dataset, our method achieved the highest accuracy 78.49\% for the second session, which is 2.38\% higher than MS-MDA, 74.34\% for the average of three sessions, which is 3.12\% higher than MS-MDA. Additionally, both F1 and Kappa scores of our model outperform the alternative methods.

To further validate the superiority of our method, we compared it with another set of state-of-the-art methods. As illustrated in Table \ref{sota_tab_cross_sub}, our method outperforms other methods in mean accuracy for both the SEED and SEED-IV datasets. The multi-source domain adaptation algorithm MFA-LR, HVF\textsubscript{2}N-DBR, and MWACN also work well in the cross-subject experiments. Our algorithm dynamically balance the model domain transferability and discriminability and achieved higher accuracy than MFA-LR (uses class-level alignment loss) and MWACN (uses association reinforcement to adapt conditional distribution). The recognition accuracy is respectively increased by 1.51\% and 3.50\% in cross-subject experiments compared to the alternative methods (HVF\textsubscript{2}N-DBR for SEED dataset and MFA-LR for SEED-IV dataset).

\subsection{Cross-Session Results}
The specific subject results of the best session in cross-session experiments are shown in Tables \ref{long_tab_cross_session}. Our algorithm, which integrates dynamic multi-source domain adaptation and class alignment techniques, achieved the highest accuracy across most subjects. On the SEED dataset, some subjects achieved the highest accuracy of 100\% in cross-session experiments. On the SEED-IV dataset, the subject achieved the highest accuracy of 95.55\% and 94.98\% in cross-session experiments. Notably, the results on the SEED dataset are significantly better than the results on the SEED-IV dataset. On the one hand, this may be because the SEED-IV dataset has more sentiment classification than the SEED dataset, which increases the difficulty of class feature recognition. On the other hand, it may be because the sample size of SEED-IV is much smaller than that of SEED. The cross-session experimental results of SEED and SEED-IV are provide in Table 
\ref{short_tab_cross_session}. Our MS-DCDA model outperformed the rest models for all four metrics.

\subsection{Ablation Studies}
\textbf{Ablation on Loss Terms} In this section, we conduct ablation experiments to investigate the different components and equilibrium factors of the model on performance. For the sake of conciseness and consistency, all ablation experiments were conducted on the SEED. The results of the ablation experiment for the model components in cross-subject and cross-session experiments are shown in Table \ref{tab11}. Compared to the accuracy of 88.75\% for full model components in cross-subject experiments, the removal of the MMD loss leads to a decrease in accuracy by 0.74\%, omitting the disc loss decreases accuracy by 2.1\%, and excluding the SCD loss results in a decrease of 4.49\%. It can be seen that SCD has the greatest performance improvement, while MMD has the smallest performance improvement. Even without using SCD, the accuracy remains higher than the baseline model by 0.13\%. Compared to the accuracy of 93.92\% for full model components in cross-session experiments, excluding the MMD loss results in a decreased accuracy of 0.51\%, removing the disc loss reduces accuracy by 1.76\%, and omitting the SCD loss leads to a decrease of 2.97\%. Similar to the cross-subject experiments, SCD shows the most substantial performance improvement, followed by disc, and MMD. The accuracy of not using SCD is still higher than that of the baseline model by 0.27\%. These results emphasize the indispensability of each loss component. 

\textbf{Ablation on Learning Strategies} Table \ref{tab12} presents the comparison of using dynamic weighting factor $\tau$ in Eq. \ref{LOSS} against several combinations of static weights. It can be seen that using dynamic weighting factor shows better performance. Using dynamic weighting factor $\tau$ promoting dynamic balance between domain alignment and class discrimination effectively avoids performance degradation caused by excessive pursuit of alignment or discriminability.
\captionsetup{font=normalsize}
\begin{table}[htbp]
\caption{Ablation Study on Loss Functions}
\setlength{\tabcolsep}{19pt} 
\small
\raggedright  
\begin{tabular}{@{}>{\hspace{6pt}}l*{3}{c}@{}} 
\toprule
\multicolumn{1}{c}{Method} &Cross-subject & Cross-session \\
\midrule
CE only &84.13$\pm$7.89&90.68$\pm$6.90\\
w/o. MMD&88.01$\pm$9.95&93.41$\pm$5.87\\
w/o. DISC &86.65$\pm$9.61&92.16$\pm$5.63\\
w/o. SCD&84.26$\pm$7.66&90.95$\pm$5.00\\
ALL&\cellcolor{LightGray}\textbf{88.75$\pm$8.88}&\cellcolor{LightGray}\textbf{93.92$\pm$8.01}\\
\bottomrule
\end{tabular}
\label{tab11}
\vspace{-0.5em}
\end{table}

\captionsetup{font=normalsize}
\begin{table}[htbp]
\caption{Ablation Study on Static vs. Dynamic Learning}
\setlength{\tabcolsep}{19.2pt} 
\small
\begin{tabular}{@{}>{\hspace{18pt}}c*{3}{c}@{}}
\toprule   
\multicolumn{1}{c}{Ratio} &Cross-subject & Cross-session \\
\midrule
1:9& 86.29 & 92.22\\
3:7& 87.27& 92.39 \\
1:1& 86.10& 93.05 \\
7:3& 87.56 & 93.39\\
9:1& 88.69& 93.32\\
$\tau$:(1-$\tau$)& \cellcolor{LightGray}\textbf{88.75}& \cellcolor{LightGray}\textbf{93.92}\\
\bottomrule
\end{tabular}
\label{tab12}
\vspace{-0.5em}
\end{table}

\textbf{Ablation on Layer Normalization}
The proposed MS- DCDA preserves the coarse-grained inter-domain alignment while improving the intra-domain and intra-class alignment through the fine-grained class-level adaptation. With the use of the SCD penalty, the increases in recognition accuracy of 4.49\% and 2.97\% are observed in the cross-subject and cross-session experiments respectively. Fine-granularity alignment encourages wider inter-class margin. As shown in Table \ref{table13}, the performance of domain alignment also depends on feature normalization, where the removal of the batch normalization layer leads to decreased performance. To improve the quality of pseudo class labels the fine-grained alignment relies on, unreliable labels are pre-filtered by an empirical threshold. The fine-grained adaptation is dynamically weighted using a dynamic coefficient, which is proportionate to the number of training iterations. By improving the domain discriminability of our model as the training proceeds, it is observed that more reliable pseudo labels for fine-grained analysis can be obtained. 

\captionsetup{font=normalsize}
\begin{table}[t]
\caption{Ablation Study on Layer Normalization (LN)}
\setlength{\tabcolsep}{1.4pt} 
\small
\raggedright  
\begin{tabular}{@{}>{\hspace{2pt}}l*{5}{c}@{}} 
            \toprule
            \multirow{2}{*}{Method} &\multicolumn{2}{c}{Cross-subject}    &  \multicolumn{2}{c}{Cross-session} \\
            \multirow{2}{*} &SEED&SEED-IV&SEED&SEED-IV\\
            \midrule
            CFE+LN & 85.51$\pm$7.99 & 70.71$\pm$9.45& 93.59$\pm$7.52  & 76.12$\pm$14.16\\
            MBC+LN  & 82.86$\pm$7.49 & 70.89$\pm$11.07& 90.06$\pm$7.21  & 77.46$\pm$12.32\\
            BOTH+LN &\cellcolor{LightGray}\textbf{88.75$\pm$8.89}&\cellcolor{LightGray}\textbf{74.34$\pm$10.62}&\cellcolor{LightGray}\textbf{93.92$\pm$8.01}&\cellcolor{LightGray}\textbf{77.53$\pm$12.13}\\
\bottomrule
\end{tabular}
\vspace{-1em}
\label{table13}
\end{table}

\captionsetup{font=normalsize}
\begin{table}[htbp]
\caption{Ablation Study on Brain Lobes}
\setlength{\tabcolsep}{2.6pt} 
\small
\raggedright  
\begin{threeparttable}
\begin{tabular}{@{}>{\hspace{2pt}}l*{5}{c}@{}} 
            \toprule
            \multirow{2}{*}{Lobes} &\multicolumn{2}{c}{Cross-subject}&\multicolumn{2}{c}{Cross-session} \\
            \multirow{2}{*} &SEED&SEED-IV&SEED&SEED-IV\\
            \midrule
            F&86.56$\pm$9.63&73.5$\pm$14.09&92.97$\pm$7.64&75.57$\pm$12.50\\
            O&85.14$\pm$8.76&66.03$\pm$9.73&91.27$\pm$7.93&73.64$\pm$12.53\\
            P&85.62$\pm$9.17&70.41$\pm$11.60&92.35$\pm$7.37&76.94$\pm$12.62\\
            T&77.93$\pm$13.19&65.45$\pm$11.81&87.78$\pm$9.37&71.31$\pm$13.59 \\    F+P&87.53$\pm$9.57&74.12$\pm$11.20&\cellcolor{LightGray}\textbf{95.22$\pm$5.71}&76.94$\pm$12.86\\
            O+T&85.55$\pm$10.84&72.10$\pm$13.22&91.91$\pm$8.02&74.86$\pm$12.07 \\
            F+P+O&86.35$\pm$8.85&72.25$\pm$12.49&92.11$\pm$7.50&77.64$\pm$13.10 \\
            F+P+T&88.33$\pm$8.78&72.55$\pm$13.26&94.57$\pm$6.10&\cellcolor{LightGray}\textbf{78.15$\pm$12.02} \\
            All&\cellcolor{LightGray}\textbf{88.75$\pm$8.89}&\cellcolor{LightGray}\textbf{74.34$\pm$10.62}&93.92$\pm$8.01&77.53$\pm$12.13\\
\bottomrule
\end{tabular}
\begin{tablenotes}[para,flushleft]
        \footnotesize
        F: Frontal; O: Occipital; P: Parietal; T: Temporal
        \end{tablenotes}
    \end{threeparttable}
    \vspace{-1.0em}
\label{table14}
\end{table}

\textbf{Ablation on Brain Lobes}
To examine impact of different brain lobes on emotion recognition, the 62 EEG electrodes are divided into four regions based on their location: front, parietal, temporal, and occidental. As shown in Table \ref{table14}, the frontal and parietal lobes appear to be the most active brain lobes while watching emotional movie clips, followed by the occipital and temporal lobes. Although only six electrodes are allocated for the temporal lobe, its high activity suggests that the expression of emotions is likely independent of the number of recognized electrode signals. The combination of frontal and parietal lobes consistently outperformed individual lobes for emotion recognition on both the SEED and SEED-IV datasets, reaffirming the important role of frontal and parietal lobes in emotional expression. In addition, as the temporal or occipital signals continue to increase and exceed a certain threshold, the recognition accuracy stops improving, and some may even appear to decrease. One possible explanation is that EEG signals measured from the occipital and temporal lobes exhibit relatively lower sensitivity to emotional states compared to those from the frontal and parietal lobes. Including the less sensitive occipital and temporal lobes may reduce the emotion recognition results. Despite of the adverse effect of the occipital and temporal lobe, the recognition accuracies remain higher in combined brain lobes compared to a single brain lobe. Our studies indicate that brain regions exhibit different sensitivities to emotional states, reaffirming the need to consider the combination of different lobes during emotion recognition. 

\subsection{Visual Analysis}
\begin{figure*}[!t]
\includegraphics[width=6.9in]{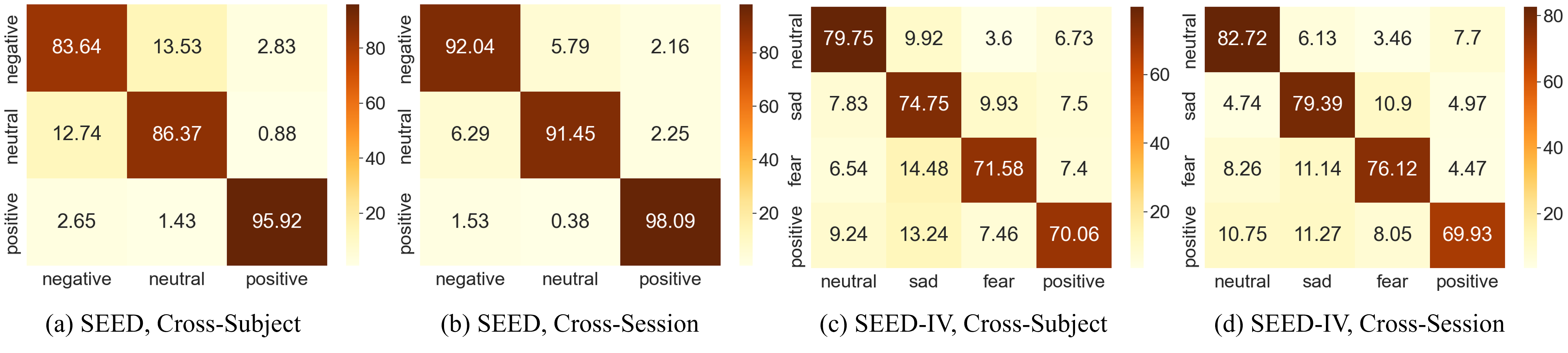}
\caption{Confusion matrices analysis of our MS-DCDA model for (a) Cross-subject experiment on SEED dataset; (b) Cross-subject experiment on SEED-IV dataset; (c) Cross-session experiment on SEED; and (d) Cross-session experiment on SEED-IV. Sensitivity are the strongest to the positive and neutral emotions in SEED and SEED-IV dataset, with slightly reduced sensitivity to the rest emotions.}
\label{Confusion}
\vspace{-1em}
\end{figure*}

\vspace{0.2cm}
\textbf{Confusion Matrix Analysis.}  The confusion matrices of cross-subject and cross-session experiments for three sessions on SEED and SEED-IV are provide in Fig. \ref{Confusion}. As can be seen in Fig. \ref{Confusion}(a) and \ref{Confusion}(b), our MS-DCDA method exhibits strongest sensitive to the positive emotion on the SEED data, with slightly reduced sensitivity in neutral and negative emtions. It is observed in Fig. \ref{Confusion}(c) and \ref{Confusion}(d) that neutral emotion is less likely being confused than fear and sadness on SEED-IV by our model, which may be explain by the fact that both fear and sadness are particularly negative and exhibit similarities in the activated electrode signals.


\begin{figure*}[!t]
\centering
\includegraphics[width=6.4in]{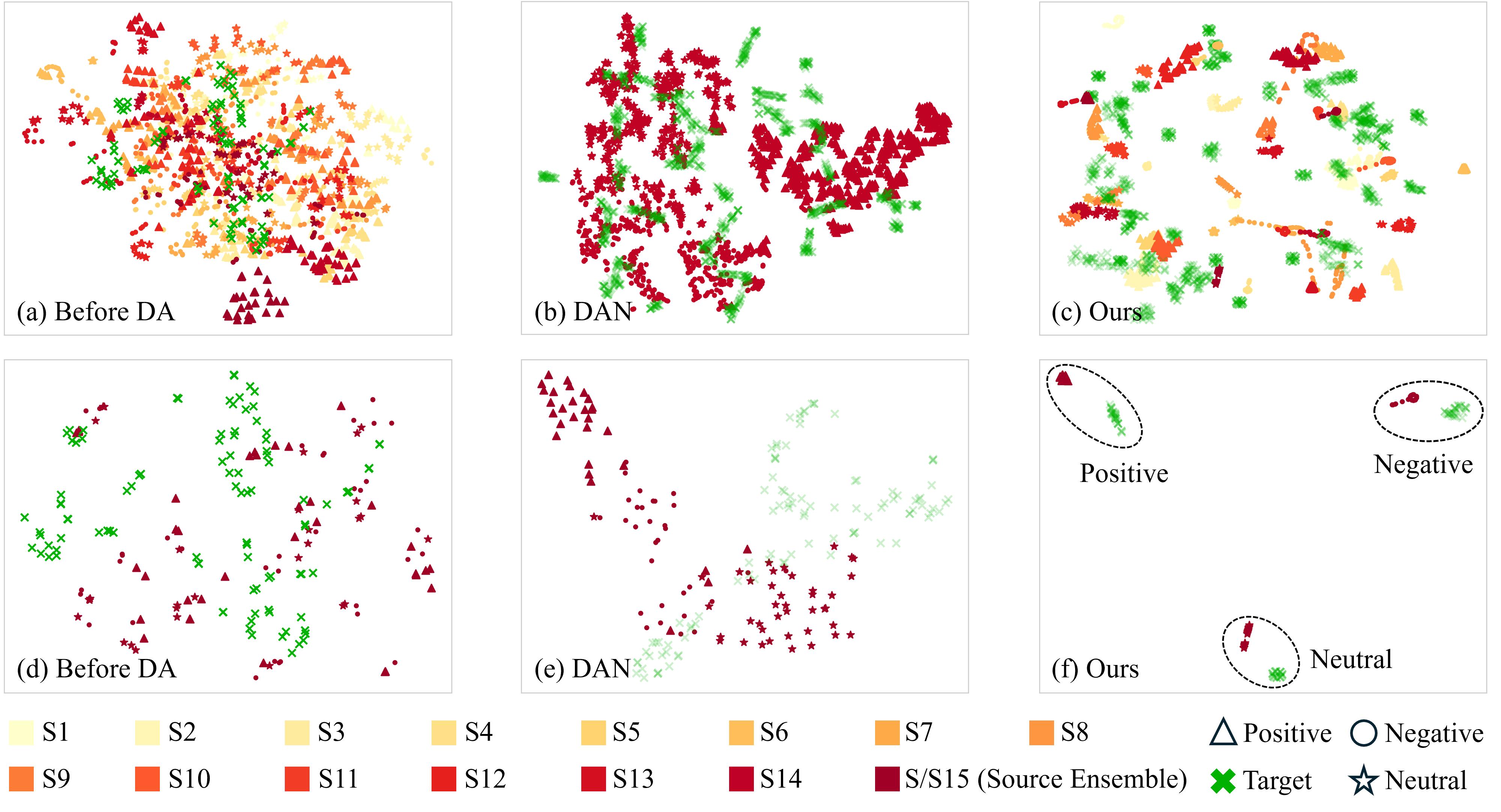}
\caption{T-SNE illustration of domain adaptation and emotion recognition on SEED dataset. (a) Distribution of 14 individual source subjects (S1-S14), the ensemble of the 14 individual sources (S15, source ensemble) and a target subject. (b) Distribution of the single source (S) and the target learned by DAN \citep{li2018cross}. (c) Distribution of the 14 individual source subjects and the target subject learned by our MS-DCDA. (d) The Distribution of the source ensemble and the target. (e) Distribution of the single source and the target learned by DAN. (f) Distribution of the source ensemble and the target learned by our MS-DCDA.}
\label{Visualization}
\vspace{-1em}
\end{figure*}

\textbf{T-SNE Analysis.} The T-SNE plots \citep{van2008visualizing} of feature distributions are provide in Fig. \ref{Visualization} for SEED dataset, which is richer in source diversity. We randomly picked 100 EEG samples from each subject for visualization to display changes in feature distribution caused by model training. As shown in Fig. \ref{Visualization}(a), the distribution of most of the samples of original features are observed to be concentrated in a singular area, with a few outliers among individual subjects. This distribution affirms that the distribution EEG features may overlap in low-level feature space. This observation becomes more clear with normalization, where a more uniform and centralized distribution appears. The feature distributions produced by the DAN model and the MS-DCDA model are shown in Fig. \ref{Visualization} (b) and Fig. \ref{Visualization} (c) respectively. Overlap between the source and target features are observed in the DAN results. On the other hand, our MS-DCDA represents each source domain and target domain feature distribution separately, showcasing similarity and overlap between the targets and each individual source domain. Fig. \ref{Visualization}(d) shows the initial distribution of combined source domain features and target domain features. We also show the feature distribution of source ensemble and target learned by DAN and our MS-DCDA respectively. While maintaining the consistency in source domain, it can be observed in Fig. \ref{Visualization}(e) that DAN does not show a clear trend of class cluttering. In contrast, our MS-DCDA model adapts T to individual sources and a complementary multi source ensemble (SE) with class-awareness. It also dynamically adjusts the weights of domain transferability and discriminability, leading to improved classification accuracies, wider inter-class margin, and higher intra-class compactness (see Fig. \ref{Visualization}(f)).

\begin{figure*}[!t]
\centering
\includegraphics[width=6.6in]{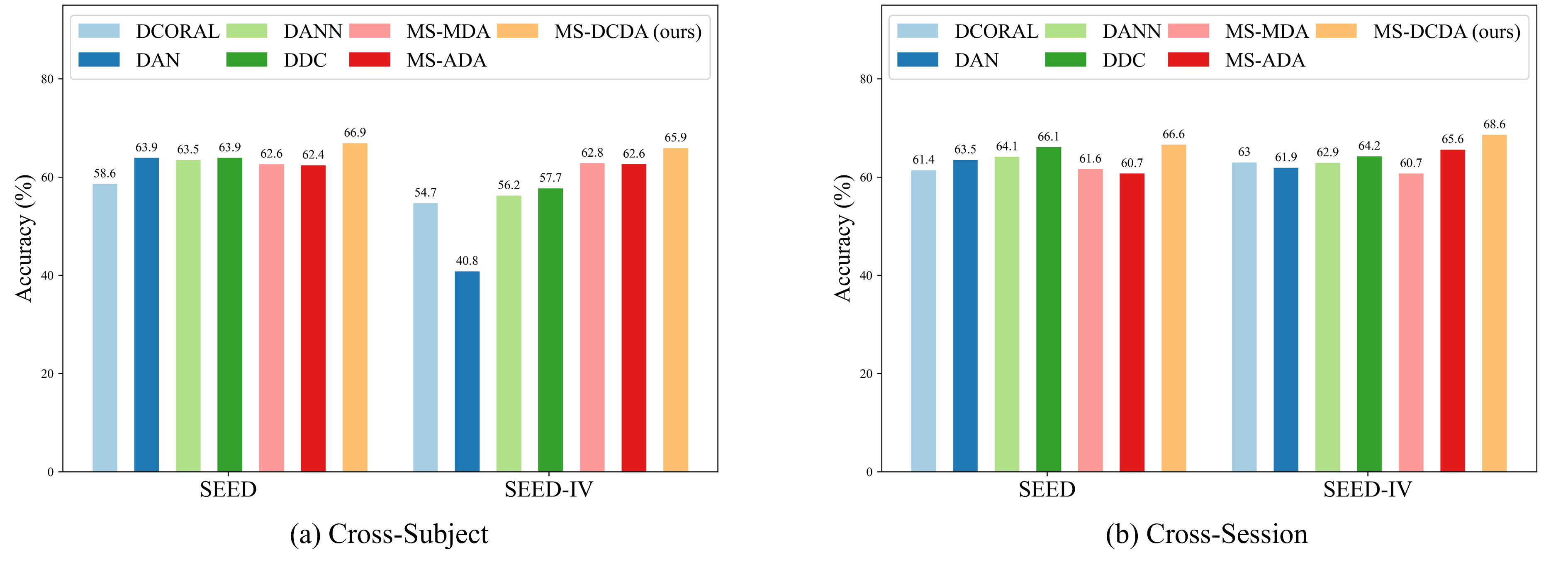}
\caption{Comparison of emotion recognition accuracies of different domain adaptation algorithms on SEED and SEED-IV datasets respectively for (a) Cross-subject experiment; and (b) Cross-session experiment. Our MS-DCDA model outperforms in all experiments four single-source methods: DCORAL \cite{sun2016deep}, DAN \cite{li2018cross}, DANN \cite{kang2020contrastive}, DDC \cite{tzeng2014deep}. The performance of our model also exceeds two multi-source methods: MS-MDA \cite{chen2021ms}, and MS-ADA \cite{she2023multisource}.}
\label{transfer}
\vspace{-2em}
\end{figure*}

\textbf{Data Transfer Studies.} The SEED and SEED-IV datasets are similar in experimental design, collection methods, data pre-processing, and analogical categories. They differ only in the number of emotion classes. Here we train our model with one dataset and test using the other as dataset transfer studies. To have the same number of emotion classes in each set, we merge the sad and fear classes in SEED-IV into a single negative class in SEED, and correspond neutral and happy in SEED-IV with the neutral and positive in SEED respectively. The results of the data transfer studies are shown in Fig. \ref{transfer}. It is observed that our MS-DCDA model outperforms alternative SS-DA and MS-DA algorithms in both cross-subject (see Fig. \ref{transfer}(a)) and cross-session experiments (see Fig. \ref{transfer}(b)), demonstrating an improved generalization ability. It is also found that the effect of data order on the performance of data transfer is minimal, especially in cross-session experiments, which may be explained by the similarity of the two datasets. The inter-class and intra-class alignment in our MS-DCDA model play important roles in extraction of fine-grained emotional features and enhancing the robustness of our model in various transfer scenarios.


\vspace{-1.5em}
\section{Discussion}\label{sec5}

The presented MS-DCDA framework demonstrated improved emotion recognition performance on both SEED and SEED-IV datasets. In cross-subject experiments, we achieved an accuracy of 90.48\% and 90.85\% for S1 and S3 on the SEED dataset, with a relatively lower accuracy of 84.98\% in S2. On the SEED-IV dataset, our MS-DCDA performs better in S2 with a leading 78.49\% accuracy, while the accuracies are slightly reduced to 72.36\% and 72.16\%  for S1 and S3 due to more complicated data dependencies in that session. The cross-session variations can result in variations in recognition accuracies if data are obtained through repetitive experiments. The robustness of the proposed framework against such variations was demonstrated through extensive cross-validation. The observation that cross-subject results appear to be less robust compared to the corresponding cross-session results may be explained by the individual subject disparities in EEG being more complicated than its non-stationary problems. As demonstrated in Tables \ref{short_tab_cross_sub}, \ref{sota_tab_cross_sub} and \ref{short_tab_cross_session}, multisource domain adaptation generally achieves superior performance than single-source methods for EEG-based emotion recognition tasks. The ablation studies on various loss functions are shown in Table \ref{tab11}, where the accuracy of our model improved from 84.13\% and 90.68\% by incorporating multi-source loss penalties. The use of pair-wise neural network branches allows adaptation of target to different source distributions through distinctive feature representations and improves the accuracy and generalization of our model. However, computational complexity may increase linearly with the number of sources. Dynamic routing among branches using a mixture-of-expert architecture may improve the recognition accuracy at a reduced training cost.


\vspace{-0.45em}
While our MS-DCDA model demonstrated improved emotion recognition performance with the use of frequency information of EEG signals, the accuracies of the emotion recognition are found to be significantly reduced without the use of DE features regardless of the recognition models. This observation reaffirms the advantage of DE features as an effective representation of EEG signals for emotion recognition. Future work may benefit from more advanced and learnable feature representations based on neural networks. End-to-end learning of EEG representation with subsequent emotion recognition models may further improve accuracy. In addition, the spatial features of the EEG data have not yet been fully explored due to the lack of spatial information of the electrodes in the current dataset. Collecting the spatial information of electrodes without significantly increasing the experiment cost may require further investigation.  Multi-modality learning that combines frequency and spatial EEG features also warrants a future study.  
Future work may involve extending the examination of EEG features in the time domain. The accuracies, generalization, and robustness of our model may also benefit from exploring the attention and gating mechanisms.



\vspace{-1.5em}

\section{Conclusion}\label{sec6}
In summary, a multi-source dynamic contrastive domain adaptation (MS-DCDA) method was introduced for EEG based emotion recognition. The proposed combines coarse- and fine-grained domain alignments and dynamically optimize the domain transferability and discriminability, leading to improved model generalization, classification accuracies, wider inter-class margin, and higher intra-class compactness. The proposed MS-DCDA outperformed both the classical and the state-of-the-art methods on both the SEED and SEED-IV datasets by a large margin. In addition, Our study also suggests greater emotional sensitivity in the frontal and parietal brain lobes, providing insights for mental health care. The broader significance of this research also lies in its potential applications in human-computer and brain-computer interfaces, allowing systems to respond more precisely to the emotional states of users through enhanced emotion recognition capabilities.

\section*{Acknowledgments} 
This work is supported by the National Natural Science Foundation of China Grants 62372371 and 61972315.

\vspace{-1em}
\bibliographystyle{elsarticle-num}
\bibliography{ref}

\end{document}